\documentclass[aps,preprint]{revtex4}%
\usepackage{amsfonts}
\usepackage{amsmath}
\usepackage{amssymb}
\usepackage{graphicx}
\usepackage{pdfpages}
\usepackage{float}
\usepackage{xcolor}
\usepackage{caption}
\usepackage{lineno,hyperref}
\usepackage{changes}%
\hypersetup{colorlinks =true, allcolors = blue}
\providecommand{\U}[1]{\protect\rule{.1in}{.1in}}

\begin{document}
\preprint{HEP/123-qed}
\title{Fermionic and Bosonic Greybody Factors as well as Quasinormal Modes for Charged Taub NUT Black Holes}
\author{Ahmad Al-Badawi}
\email{ahmadbadawi@ahu.edu.jo}
\affiliation{Department of Physics, Al-Hussein Bin Talal University, P. O. Box: 20, 71111,
Ma'an, Jordan.}
\author{Sara Kanzi}
\email{sara.kanzi678@gmail.com}
\affiliation{Faculty of Engineering, Final International University,
Kyrenia, North Cyprus via Mersin 10, Turkey}
\author{}
\affiliation{}
\author{\.{I}zzet Sakall{\i}}
\email{izzet.sakalli@emu.edu.tr}
\affiliation{Physics Department, Eastern Mediterranean
University, Famagusta, North Cyprus via Mersin 10, Turkey.}
\author{}
\affiliation{}

\begin{abstract}
The paper studies the spinorial wave equations, namely the Dirac and the Klein–Gordon equations, as well as the greybody radiations and  quasinormal modes (QNMs) of the charged Taub NUT black hole (CTNBH). To obtain fermionic greybody factors (GFs) and QNMs, we study the charged fermions by employing the Dirac equation. To this end, we use a null tetrad in the Newman-Penrose formalism. Then, we separate the Dirac equation into radial and angular sets. Using the obtained radial equations, we convert them into the typical one dimensional Schr\"{o}dinger-like wave equations with the aid of tortoise coordinate and derive the effective potentials. For bosonic GFs and QNMs, we study the Klein-Gordon equation in the CTNBH geometry and obtain the radial equation. We then derive the effective potential and investigate the effect of NUT parameter on it. We show that while the fermionic QNMs and GFs individually increase with the increasing NUT parameter, the increase of bosonic GFs with increasing NUT parameter is overwhelmingly greater than that of the bosonic QNMs.
\end{abstract}
\volumeyear{ }
\eid{ }
\date{\today}
\received{}

\maketitle
\tableofcontents
\section{Introduction}
If we look at the history of Taub-NUT (TN) spacetime, we see that it dates back to 1951 when Abraham Haskel Taub \cite{is1} discovered the underlying Taub space, which was later expanded to a wider manifold by Ezra T. Newman, Louis A. Tamburino, and Theodore W. J. Unti \cite{is2}, whose initials constitute the  ``NUT" of the TN spacetimes. The metric of TN is yet another exact solution to the Einstein's equations. It might be regarded as a first effort for determining a spinning black hole (BH) metric. It is also occasionally utilized in general relativity-based cosmological models that are homogenous yet anisotropic. On the other hand, in 1963, when Roy Kerr \cite{is3} introduced the Kerr metric for rotating BHs, he came up with a 4-parameter solution, one of which was the mass of the central body and the other was its angular momentum. The NUT-parameter or the so-called NUT charge \cite{is2} was one of the two remaining parameters, which was eliminated from his solution because Kerr believed that it was nonphysical since it made the metric non-asymptotically flat. However, other researchers interpret it either as a gravomagnetic monopole \cite{is4} parameter of the central mass or a twisting property of the surrounding spacetime \cite{is5,is6}. So, when the  NUT charge is viewed as gravitational magnetic charge, the solution can be interpreted as a sort of gravitational dyon and its Euclidean continuation (for certain values of the mass and NUT charge) is nothing but a Kaluza-Klein monopole \cite{is7}. The TN solutions also have a singular one-dimensional string, dubbed the Misner string, that is similar to the Dirac string \cite{is8}. Since the singularity of the Misner
string is a coordinate singularity, we can not describe the manifold using one coordinate
system. Moreover, Dumitru Astefanesei, Robert B. Mann, and  Eugen Radu \cite{is9} showed that locally asymptotically AdS spacetimes with nonzero NUT charge do not obey the conventional area-entropy relationship. This finding is a result of the first law of thermodynamics, not the existence of closed timelike curves or the elimination of Misner string singularities. The charged TN solutions also introduce a fascinating family of solutions known as the Israel-Wilson-Perjés solutions, which have crucial qualities in terms of supersymmetry and duality \cite{is10}. Today, there exist many researchers on  BHs with NUT parameters \cite{isS10,isS11}, including holographic complexity of charged
TN-AdS BH \cite{is11}, gravitational lensing \cite{is12}, particle acceleration \cite{is13}, and holography \cite{is14}. 

The gravitational perturbations of a static BH geometry were studied
several decades ago \cite{is15,is16}. In the sequel, the same idea was applied to the studies of other types perturbations of
various (stationary, lower dimensional, higher dimensional etc.) BHs caused by fields of a different nature like spin-$0,1,1/2$.. fields \cite{is17}. Those works led to the developments of QNM and GF methods \cite{is18,is19,IS19,IS20,IS21}.
Ultimately, scientists obtained important information regarding the stability and thermal radiation of BHs against
perturbations occuring in the exterior region. 

In this study, we consider the charged TNBH (CTNBH) presented by Dieter R. Brill
\cite{is20} and which was generalized in \cite{is21,is22}. The considered CTNBH solution depends on three parameters: mass $M$, NUT parameter $l$, and charge $q$. We mainly study the spinorial wave dynamics in the geometry of CTNBH with the aid of the Klein-Gordon and Dirac wave equations \cite{is17,is23}. After having the master equations, we compute the GFs and QNMs of the CTNBH. The physical interpretations of the results obtained will be supported by graphics and tables.
    
Our paper is organized as follows: in Sect. \ref{sec2},  we briefly
review the charged CTNBH and present its some of physical features. Next, we study the spinorial wave equations, namely the Dirac and Klein-Gordon equations, in the background of the CTNBH. Section \ref{sec3} computes the GFs of the CTNBH spacetime for both fermions and bosons. Section \ref{sec4} is devoted to the computations of the QNMs of the  fermionic and bosonic fields propagating in the geometry of the CTNBH. We draw our conclusions in Sect. \ref{sec5}.

\section{Spinorial wave equations} \label{sec2}
The CTNBH is a solution to the Einstein-Maxwell theory. Its metric is given by \cite{is20}
\begin{equation}
ds^{2}=f(r)\left[ dt-2l\cos \theta d\phi \right] ^{2}-\frac{1}{f(r)}%
dr^{2}-\left( r^{2}+l^{2}\right) \left( d\theta ^{2}+\sin ^{2}\theta d\phi
^{2}\right), \label{isq1}
\end{equation}
where the metric function reads
\begin{equation}
f(r)=1-\frac{2\left( Mr+l^{2}\right) +q^{2}}{r^{2}+l^{2}},\label{isq2}
\end{equation}%
and the electromagnetic potential is
\begin{equation}
\mathbf{A}=\frac{q r}{r^{2}+l^{2}}(d t-2l \cos \theta d \phi), \label{isq3}
\end{equation}
where $M$, $l$, and $q$ denote the mass, the NUT parameter, and the electric parameter of the BH, respectively. It can be easily seen that for $l=0$, one recovers the Reissner–Nordstr\"{o}m solution. Due to the absence of a spacetime singularity in the CTNBH solution (\ref{isq1}), this BH is also known as a regular solution of the Reissner-Nordstr\"{o}m BH. 
\subsection{Dirac Equation}
Within the framework of Newman-Penrose formalism \cite{is24}, the Dirac equations that govern the behavior of a spin-$\frac{1}{2}$ charged particle are given by \cite{is17}  
\[
\left( D+iel^{\mu }A_{\mu }+\epsilon -\rho \right)
F_{1}+\left( \bar{\delta}+ie\overline{m}^{\mu }A_{\mu
}+\pi -\alpha \right) F_{2}=i\mu _{\ast }G_{1},
\]%
\[
\left( \Delta+ien^{\mu }A_{\mu }+\mu -\gamma \right)
F_{2}+\left(\delta+iem^{\mu }A_{\mu }+\beta -\tau \right)
F_{1}=i\mu _{\ast }G_{2},
\]%
\[
\left( D+iel^{\mu }A_{\mu }+\overline{\epsilon }-%
\overline{\rho }\right) G_{2}-\left( \delta +ie m^{\mu
}A_{\mu }+\overline{\pi }-\overline{\alpha }\right) G_{1}=i\mu _{\ast }F_{2},
\]%
\begin{equation}
\left( \Delta+ien^{\mu }A_{\mu }+\overline{\mu }-\overline{%
\gamma }\right) G_{1}-\left(\bar{\delta} +ie\overline{%
m}^{\mu }A_{\mu }+\overline{\beta }-\overline{\overline{\tau }}\right)
G_{2}=i\mu _{\ast }F_{1},\label{isq4}
\end{equation}

where $\mu _{0}$ and $e$ denote the mass and charge of the Dirac particle, respectively. The $D, \Delta$ are the directional derivatives and $l^{\mu}, n^{\mu},m^{\mu}$ are the the dual null co-tetrads (see the appendix). To solve the Dirac equation \eqref{isq4}, we assume
\begin{equation}
F_{1}=F_{1}\left( r,\theta \right) e^{i\left( kt+n\phi \right) },  \nonumber
\end{equation}%
\[
F_{2}=F_{2}\left( r,\theta \right) e^{i\left( kt+n\phi \right) },
\]%
\[
G_{1}=G_{1}\left( r,\theta \right) e^{i\left( kt+n\phi \right) },
\]%
\begin{equation}
G_{2}=G_{2}\left( r,\theta \right) e^{i\left( kt+n\phi \right) }, \label{isq5}
\end{equation}%
where $k$ is the frequency of the incoming wave and $n$ represents the azimuthal quantum number. Substituting the appropriate non-zero spin coefficients (see Appendix \ref{A1N}-\ref{A4N}) together with the spinors (\ref{isq5}) into the Dirac equation  (\ref{isq4}), we obtain \[
\sqrt{r^{2}+l^{2}}\left( \mathcal{D}-\frac{il}{2\left( r^{2}+l^{2}\right) }%
\right) F_{1}+\frac{1}{\sqrt{2}}\mathit{L}F_{2}=i\mu _{0}\sqrt{r^{2}+l^{2}}%
G_{1},
\]%
\[
\frac{\sqrt{f}}{2}\sqrt{r^{2}+l^{2}}\left( \mathcal{D}^{\dag }-\frac{il}{%
2\left( r^{2}+l^{2}\right) }\right) \sqrt{f}F_{2}-\frac{1}{\sqrt{2}}\mathit{L%
}^{\dag }F_{1}=-i\mu _{0}\sqrt{r^{2}+l^{2}}G_{2},
\]%
\[
\sqrt{r^{2}+l^{2}}\left( \mathcal{D}+\frac{il}{2\left( r^{2}+l^{2}\right) }%
\right) G_{2}-\frac{1}{\sqrt{2}}\mathit{L}^{\dag }G_{1}=i\mu _{0}\sqrt{%
r^{2}+l^{2}}F_{2},
\]%
\begin{equation}
\frac{\sqrt{f}}{2}\sqrt{r^{2}+l^{2}}\left( \mathcal{D}^{\dag }+\frac{il}{%
2\left( r^{2}+l^{2}\right) }\right) \sqrt{f}G_{1}+\frac{1}{\sqrt{2}}\mathit{L%
}G_{2}=-i\mu _{0}\sqrt{r^{2}+l^{2}}F_{1}, \label{isq6} 
\end{equation}
in which $\mu _{0}=\sqrt{2}\mu _{\ast }$ and 
\[
\mathcal{D}=\frac{\partial }{\partial r}+\frac{r}{r^{2}+l^{2}}+\frac{i}{f}%
\left( k+\frac{qer}{r^{2}+l^{2}}\right) ,
\]%
\[
\mathcal{D}^{\dag }=\frac{\partial }{\partial r}+\frac{r}{r^{2}+l^{2}}+\frac{%
f^{\prime }}{2f}-\frac{i}{f}\left( k+\frac{qer}{r^{2}+l^{2}}\right) 
\]%
\[
\mathit{L}=\partial _{\theta }+\left( \frac{1}{2}+2lk\right) \cot \theta +\frac{n}{\sin \theta },
\]%
\begin{equation}
\mathit{L}^{\dag }=\partial _{\theta }+\left( \frac{1}{2}-2lk\right) \cot \theta -\frac{n}{\sin \theta }. \label{isq7}
\end{equation}
To separate the above equations (\ref{isq6}) into radial and angular equations, we set 
\begin{eqnarray}
F_{1} &=&R_{+1/2}\left( r\right) A_{1}\left( \theta \right) , \nonumber \\
F_{2} &=&R_{-1/2}\left( r\right) A_{2}\left( \theta \right) ,  \nonumber \\
G_{1} &=&R_{-1/2}\left( r\right) A_{1}\left( \theta \right) ,  \nonumber \\
G_{2} &=&R_{+1/2}\left( r\right) A_{2}\left( \theta \right) . \label{isq8}  
\end{eqnarray}%
With this assumption, Eqs. \eqref{isq6} become
\begin{eqnarray}
\sqrt{2\left( r^{2}+l^{2}\right) }\mathcal{D}R_{+1/2} &=&\left( \lambda +i\mu
_{0}\sqrt{2\left( r^{2}+l^{2}\right) }\right) R_{-1/2},\label{isq9} \\
\sqrt{2\left( r^{2}+l^{2}\right) }\frac{f}{2}\mathcal{D}^{\dag }R_{-1/2}
&=&\left( \lambda -i\mu _{0}\sqrt{2\left( r^{2}+l^{2}\right) }\right)
R_{+1/2} \label{isq10}  
\end{eqnarray}
\begin{equation}
\mathit{L}A_{2}=\lambda A_{1},
\end{equation} \label{isq11} 
\begin{equation}
\mathit{L}^{\dag }A_{1}=-\lambda A_{2}. \label{isq12}
\end{equation}
\bigskip The radial equations \eqref{isq9} and \eqref{isq10} can be rearranged as
\begin{equation}
\sqrt{r^{2}+l^{2}}\left( \frac{d}{dr}+\frac{r}{r^{2}+l^{2}}+\frac{i}{f}%
\left( k+\frac{qer}{r^{2}+l^{2}}\right) \right) R_{+1/2}=\left( \lambda
-i\mu _{0}\sqrt{r^{2}+l^{2}}\right) R_{-1/2},  \label{isq13}
\end{equation}%
\begin{equation}
\frac{f}{2}\sqrt{r^{2}+l^{2}}\left( \frac{d}{dr}+\frac{r}{r^{2}+l^{2}}+\frac{%
f^{\prime }}{2f}-\frac{i}{f}\left( k+\frac{qer}{r^{2}+l^{2}}\right) \right)
R_{-1/2}=\left( \lambda +i\mu _{0}\sqrt{r^{2}+l^{2}}\right) R_{+1/2}.
\label{isq14}
\end{equation}

To transform the radial equations to the form of one-dimensional Schr\"{o}dinger  wave equations, we perform first following transformations:
\begin{equation}
P_{+1/2}=\sqrt{r^{2}+l^{2}}R_{+1/2},\qquad P_{-1/2}=\sqrt{\frac{f}{2}}\sqrt{%
r^{2}+l^{2}}R_{-1/2}. \label{isq15}
\end{equation}

Hence, Eqs. \eqref{isq13} and \eqref{isq14} recast in 
\begin{equation}
\left( \frac{d}{dr}+ik\frac{\Omega }{f}\right) P_{+1/2}=\sqrt{\frac{2}{f}}%
\left( \frac{\lambda }{\sqrt{r^{2}+l^{2}}}-i\mu _{0}\right) P_{-1/2}, \label{isq16}
\end{equation}
\begin{equation}
\left( \frac{d}{dr}-ik\frac{\Omega }{f}\right) P_{-1/2}=\sqrt{\frac{2}{f}}%
\left( \frac{\lambda }{\sqrt{r^{2}+l^{2}}}+i\mu _{0}\right) P_{+1/2}, \label{isq17}
\end{equation}
where $\Omega =\left( 1+\frac{qer}{\left( r^{2}+l^{2}\right) k}\right).$ 
Assuming 
\begin{equation}
\frac{du}{dr}=\frac{\Omega }{f},\label{isq18}
\end{equation}
equations \eqref{isq16} and \eqref{isq17} become
\begin{equation}
\left( \frac{d}{du}+ik\right) P_{+1/2}=\frac{\sqrt{2f}}{\Omega }\left( \frac{%
\lambda }{\sqrt{r^{2}+l^{2}}}-i\mu _{0}\right) P_{-1/2},\label{isq19}
\end{equation}%
\begin{equation}
\left( \frac{d}{du}+ik\right) P_{-1/2}=\frac{\sqrt{2f}}{\Omega }\left( \frac{%
\lambda }{\sqrt{r^{2}+l^{2}}}+i\mu _{0}\right) P_{+1/2}. \label{isq20}
\end{equation}%
Now, let us apply another transformation:
\begin{equation}
P_{+1/2}=\phi _{1}\exp \left( \frac{i}{2}\tan ^{-1}\left( \frac{\mu _{0}}{%
\lambda }\sqrt{r^{2}+l^{2}}\right) \right),\quad  P_{-1/2}=\phi _{1}\exp \left( 
\frac{-i}{2}\tan ^{-1}\left( \frac{\mu _{0}}{\lambda }\sqrt{r^{2}+l^{2}}%
\right) \right), \label{isq21}
\end{equation}

and change the variable from $u$ to $\widehat{r}=u+\frac{1}{2k}\tan
^{-1}\left( \frac{\mu _{0}}{\lambda }\sqrt{r^{2}+l^{2}}\right) $, thus we get%
\begin{equation}
\frac{d\phi _{1}}{d\widehat{r}}+ik\phi _{1}=W\phi _{2},\label{isq22}
\end{equation}%
\begin{equation}
\frac{d\phi _{2}}{d\widehat{r}}+ik\phi _{2}=W\phi _{1}, \label{isq23}
\end{equation}
where 
\begin{equation}
W=\frac{2k\sqrt{f}\left( \frac{\lambda ^{2}}{r^{2}+l^{2}}+\mu
_{0}^{2}\right) ^{3/2}}{2\Omega k\left( \frac{\lambda ^{2}}{r^{2}+l^{2}}+\mu
_{0}^{2}\right) +f\lambda \mu _{0}}.    \label{isq24} \end{equation}

Finally,  we define $2\phi _{1}=\Psi _{1}+\Psi _{2},\quad 2\phi _{2}=\Psi
_{1}-\Psi _{2}$ and then Eqs. \eqref{isq22} and \eqref{isq23} become
\begin{equation}
\frac{d^{2}\phi _{1}}{d\widehat{r}}+k^{2}\Psi _{1}=V_{+}\Psi _{2},\label{isq25}
\end{equation}
\begin{equation}
\frac{d^{2}\phi _{2}}{d\widehat{r}}+k^{2}\Psi _{2}=V_{-}\Psi _{1}, \label{isq26}
\end{equation}
where the effective potentials are obtained as follows 
\begin{equation}
V_{\pm }=W^{2}\pm \frac{dW}{d\widehat{r}}.  \label{isq27}
\end{equation}
The above potentials are normally quite elongated, however they get simpler form for the massless fermions (neutrinos) by setting $\mu _{0}=0$ in \eqref{isq27}:
\begin{equation}V_{\pm }=\frac{f\lambda ^{2}}{D^{2}}\pm \frac{\sqrt{f}\lambda }{D^{2}}\left[
\left( r-M\right) \right] \mp \frac{f^{3/2}\lambda }{D^{3}}\left[ 2r\left( 1+
\frac{qer}{\left( r^{2}+l^{2}\right) k}\right) -r^{2}\left( \frac{qer}{
\left( r^{2}+l^{2}\right) ^{2}k}\right) \right] , \label{isq28}
\end{equation}
where%
\begin{equation}
D=r^{2}\left( 1+\frac{qer}{\left( r^{2}+l^{2}\right) k}\right). \label{isq29}
\end{equation}
\begin{figure}
    \centering
    {{\includegraphics[width=8cm]{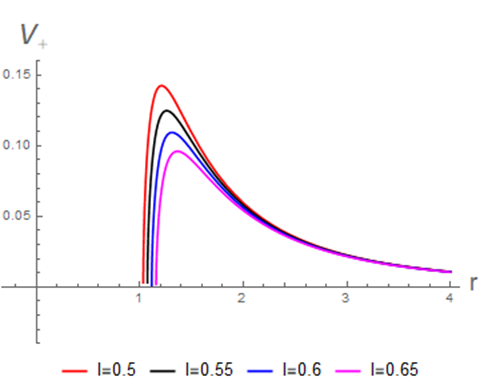}}}\qquad
{{\includegraphics[width=7cm]{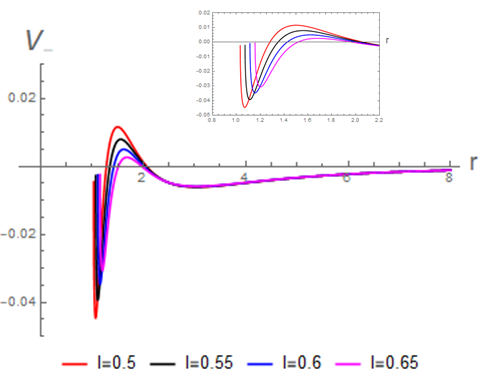} }}
    \caption{The behaviors of potentials (\ref{isq28})
for various values of the NUT parameter $l$ ( $V_{+}$ and $V_{-}$ are the left and right plots, respectively).  The physical parameters are chosen as $\lambda =  1, k = 0.2, q =0.1, e = 0.5$, and $M = 0.4$.}
    \label{fig1}
\end{figure}
To examine NUT's effect on the potentials, we plot the potentials against distance. Figure \ref{fig1} represents the behavior of the potentials \eqref{isq28} for some specific values of NUT parameter. It is obvious from Fig. \ref{fig1} that higher NUT parameters lead to smaller potentials, so that their main effects are to attenuate peaks and modify the potential barriers.
\subsection{Klein-Gordon Equation}

\begin{figure}[h]
\centering
\includegraphics[scale=.5]{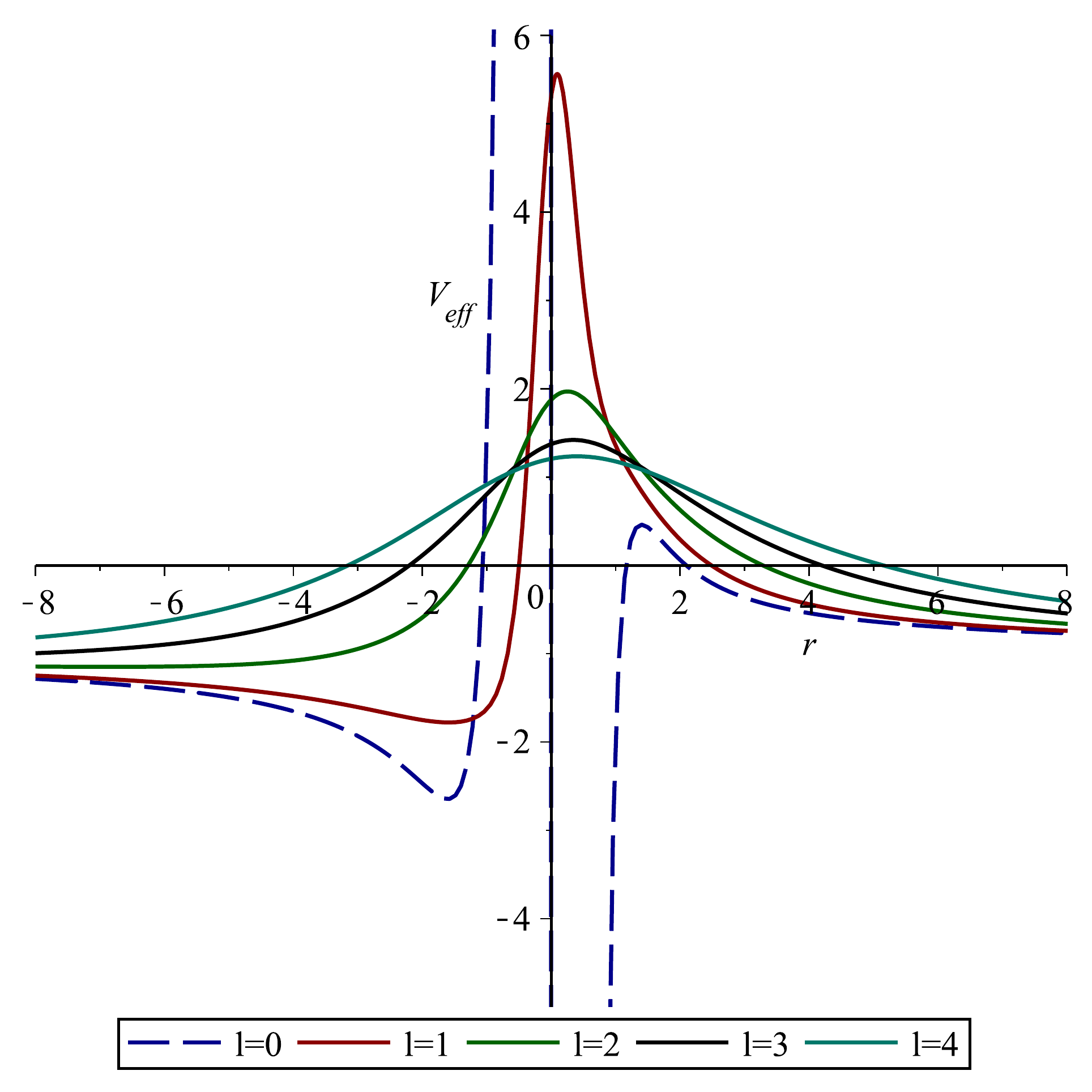} \caption{Plots of $V_{eff}$ \eqref{isq36}
for various values of the NUT parameter $l$. Here, the chosen physical parameters are $\lambda =  2, q =\tilde{q}= 0.5$, and $M =\mu= 1$.}
\label{Figure3}%
\end{figure}

In this section, the Klein-Gordon equation \cite{is17} of a scalar field with mass $\mu_{0}$ is examined in the CTNBH geometry. We shall reduce the radial wave equation to a one-dimensional Schr\"{o}dinger like wave equation and derive the effective potential. The massive charged Klein-Gordon equation is given by \cite{is25}
\begin{equation}
\frac{1}{\sqrt{-g}}\left(\partial_{\mu}-i\tilde{q}A_{\mu}\right)[\sqrt{-g}g^{\mu\nu}\left(\partial_{\nu}-i\tilde{q}A_{\nu}\right)]\Phi=\mu_{0}^{2}\Phi , \label{isq30}   
\end{equation}
by which $\mu$ and $\tilde{q}$ are the mass and charge of the scalar field $\Phi$, respectively. By utilizing the following ansatz
\begin{equation}
    \Phi\left(t,r,\theta,\phi\right)=e^{-i\omega t}e^{im\phi}R(r)Y(\theta), \label{isq31}
\end{equation}
one can derive the angular equation as 
\begin{equation}
    \left[\frac{1}{sin\theta}\frac{d}{d\theta}\left(sin\theta\frac{dY}{d\theta}\right)-\left(2l\omega cot\theta-\frac{m}{sin\theta}\right)^{2}-\lambda\right]Y(\theta)=0, \label{isq32}
\end{equation}
which is nothing but the Legendre differential equation with the eigenvalue of $\lambda=\tilde{l}(\tilde{l}+1)$. Besides, the radial equation reads
\begin{equation}
    \frac{d}{dr}\left[(r^2+l^2)f(r)\frac{dR}{dr}\right]+\left(\frac{\left(\omega(r^2+l^2)+\tilde{q}qr\right)^{2}}{f(r)(r^2+l^2)}+\lambda+\mu_{0}^{2}(r^2+l^2)\right)R(r)=0. \label{isq33}
\end{equation}
To reach a Schr\"{o}dinger-like wave equation from Eq. \eqref{isq33}, two more steps are required. First, we apply the transformation of $R=\frac{U}{\sqrt(r^2+l^2)}$ and in the sequel use the tortoise coordinate $r_{\star}=\int\frac{dr}{f(r)}$. After straightforward computations, Eq. \eqref{isq33} becomes
\begin{equation}
    \frac{d^{2}U}{dr_{\star}^2}+(\tilde{\omega^2}-V_{eff})U=0,\label{isq34}
\end{equation}
in which 
\begin{equation}
    \tilde{\omega}=\omega^{2}+\frac{2\omega\tilde{q}qr}{(r^2+l^2)}, \label{isq35}
\end{equation}
and
\begin{equation}
    V_{eff}=\frac{f(r)}{(r^2+l^2)}\left[f^{\prime}(r)+\frac{l^2f(r)}{(r^2+l^2)}-\frac{\tilde{q}^{2}q^{2}r^2}{f(r)(r^2+l^2)}-\lambda-\mu_{0}^{2}(r^2+l^2)\right].\label{isq36}
\end{equation}
The way of behaving of the effective potential \eqref{isq36} under the varying NUT parameter $l$ is depicted in Fig. \ref{Figure3}. As can be seen from  Fig. \ref{Figure3}, the dissimilarity with and without of the existence (specially for $l=1$) of the NUT parameter is significant. Moreover, the maximum peak at $l=1$ involves a subsidence when increasing the NUT parameter.

\section{GF\lowercase{s} of charged CTNBH} \label{sec3}
In this section, we shall investigate the bosonic greybody radiation of the CTNBH. To this end, the lower bound semianalytic approach \cite{GF1,GF2,GF3,GF4,GF5,GF6,GFs6} is employed. We start our computations with the definition of GF \cite{GF1,GFL6}: 
\begin{equation}
\sigma_{\ell}\left(  \omega\right)  \geq\sec h^{2}\left(  \int_{-\infty
}^{+\infty}\wp dr_{\ast}\right)  ,\label{isq37}%
\end{equation}
where
\begin{equation}
\wp=\frac{\sqrt{\left(  h%
%TCIMACRO{\U{b4}}%
%BeginExpansion
\acute{}%
%EndExpansion
\right)  ^{2}+\left(  \omega^{2}-V_{eff}-h^{2}\right)  ^{2}}}{2h},\label{isq38}%
\end{equation}

Applying the following conditions: 1) $h\left(  r_{\ast}\right)
> 0$ and 2) $h\left(  -\infty\right)  =h\left(  \infty\right)  =\omega$, and if one simply sets $h=\omega$ \cite{GF2}, thus the GF formula (\ref{isq37}) reduces to
\begin{equation}
\sigma_{\ell}\left(  \omega\right)  \geq\sec h^{2}\left(  \int_{-\infty
}^{+\infty}\frac{V_{eff}}{2\omega}dr_{\ast}\right)  \text{.}\label{isq39}%
\end{equation}
\subsection{Bosonic GF\lowercase{s}}
\begin{figure}[h]
\centering
\includegraphics[scale=.5]{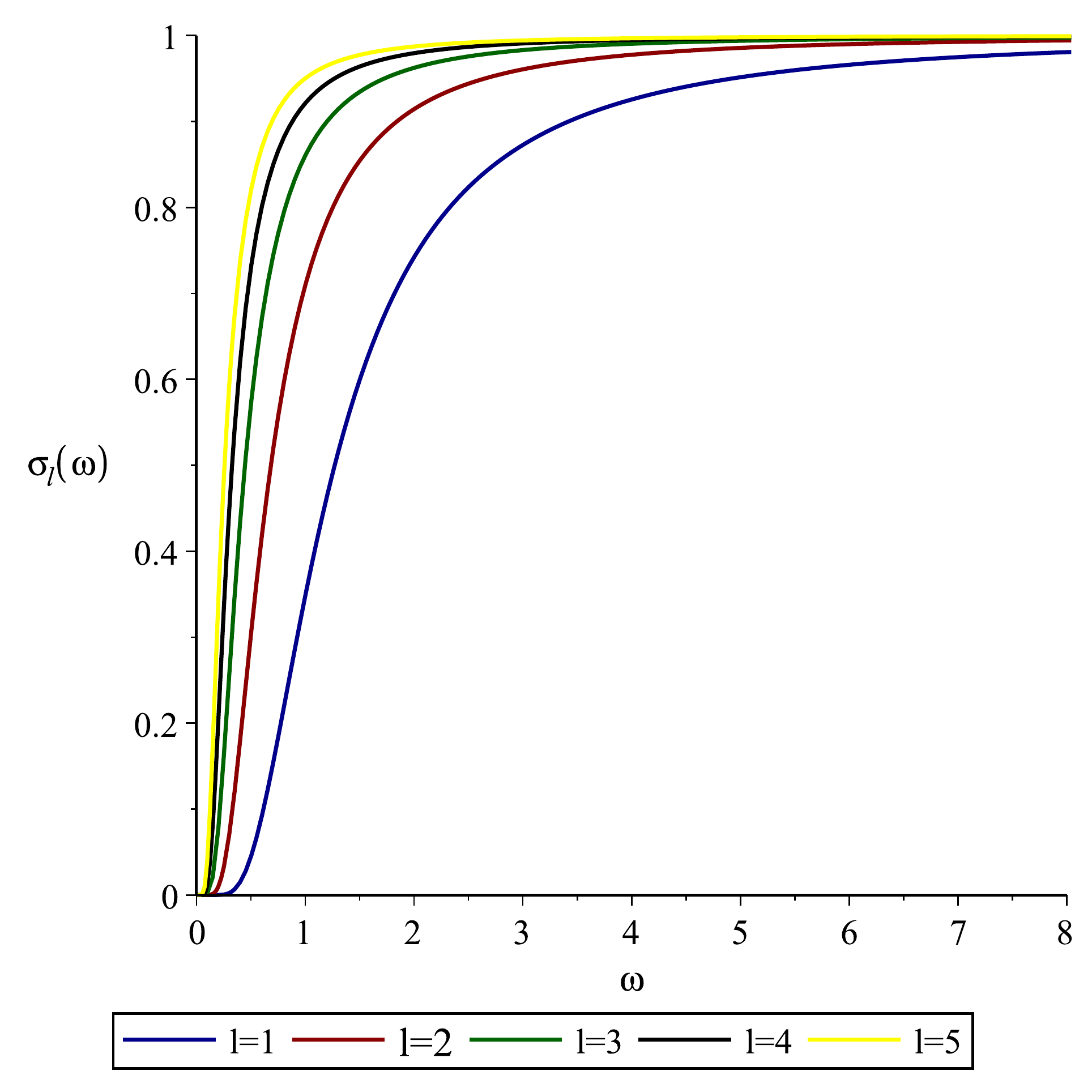} \caption{Graph of $\sigma_{\ell}$ (\ref{isq40})
for various values of the NUT parameter $l$.  Here, the chosen physical parameters are $\lambda =  2, q =\tilde{q}= 0.5$, and $M = 1$.}
\label{Figure4}%
\end{figure}

To define the GFs of scalar particles, we should consider the regarded effective potential. Therefore,
substituting Eq. \eqref{isq36} for the massless particles $(\mu_{0}=0)$ in Eq. \eqref{isq39} and integrating it with respect to $r$, one can obtain the GFs of bosons emitted from the CTNBH as follows
   
\begin{multline}
\sigma_{\ell}\left(  \omega\right)  \geq\sec h^{2}\left(  \frac{1}{2\omega
}\left[  -\frac{r_{h}}{4\left(  r_{h}^{2}+l^{2}\right)  }+\frac{Mr_{h}}{2l^{2}(l^{2}+r_{h}^{2})}-\frac{\arctan(\frac{r_{h}}{l})}{4l}\left(1-\frac{2M}{l^2}+\frac{3Q^2}{2l^2}+4\lambda\right)\right.  \right.
\\
\left. - \frac{2l^2r_{h}+q^2r_{h}+4M-2Ml^2+4l^2+2q^2}{4(r_{h}^2+l^2)}
-\frac{3rq^{2}}{8l^2(r_{h}^2+l^2)}-\right.  \\
\left.  \left.  -\frac{\tilde{q}^{2}q^{2}}{r_{h}}\left(1+\frac{M}{r_{h}}+\frac{4M^2+q^2}{3r_{h}^2}\right)  \right] \right). \label{isq40}
\end{multline}
After inserting $r_{h}=M\pm\sqrt{M^{2}+r^{2}+q^{2}}$ into the  Eq. \eqref{isq40} and, we get the scalar GFs  of the CTNBH in terms of the BH parameters ($M,q$). The behavior of the bosonic GFs of the CTNBH is depicted Fig. \ref{Figure4}, which shows the effect of the NUT parameter $l$ on the GF.

\subsection{Fermionics GF\lowercase{s}}
\begin{figure}[h]
\centering
\includegraphics[scale=.5]{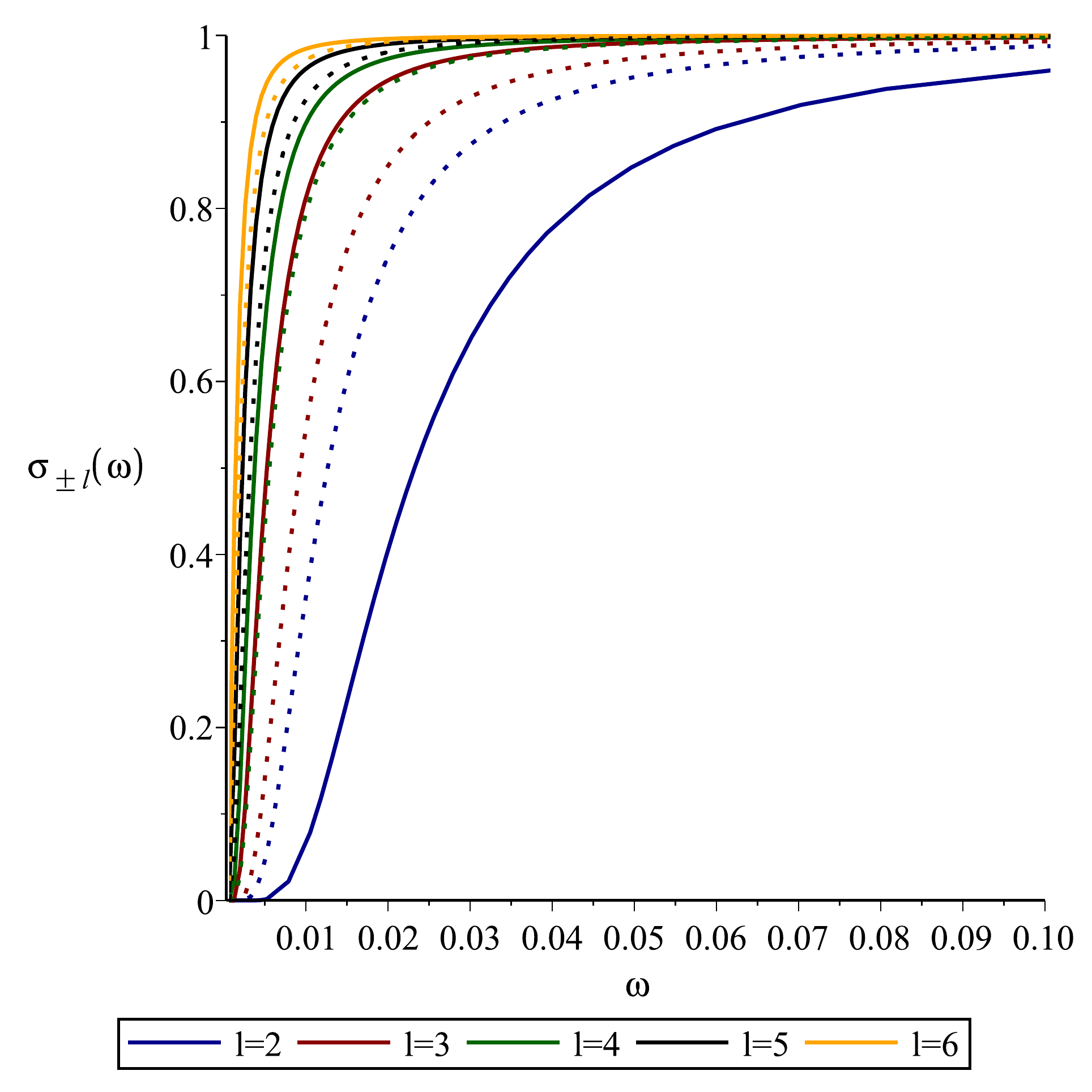} \caption{Graph of $\sigma_{\pm\ell}(\omega)$ \eqref{isq42}
for various values of the NUT parameter $l$. While the solid lines indicate the spin-$(+\frac{1}{2})$, dotted lines represent spin-$(-\frac{1}{2})$ particles.  Here, the physical parameters are chosen as $\lambda = -1.5, k = 0.2, q =0.1, e = 0.5$, and $M = 1$.}
\label{Figure5}%
\end{figure}

 In this part, we shall investigate the fermionic GFs on the CTNBH. For this purpose, we consider the effective potential \eqref{isq28} of the fermions and use it in Eq. \eqref{isq39}. Thus, we get the following integral expression
\begin{equation}
    \sigma_{\pm l}(\omega)\geq\sec h^{2}\left[\frac{\lambda}{2\omega}\left(\int_{r_{h}
}^{+\infty}\left(\frac{\lambda}{D^2}\pm\frac{(r-M)}{D^2\sqrt{f}}\mp\frac{\sqrt{f}}{D^3}\left[\frac{2D}{r}-r^{2}\left(\frac{qer}{(r^2+l^2)^2k}\right)\right]\right)dr\right)\right]. \label{isq41}
\end{equation}
Normally the complicated nature of the above integral did not allow the analytical solution. To overcome this issue, the integrand of Eq. \eqref{isq41} is expended to the series under the low charge approximation. After making some straightforward computations, one can find out the following result

\begin{multline}
\sigma_{\pm\ell}(\omega)\geq\sim\sec h^{2}\left(  \frac{\lambda}{2\omega
}\left[  \frac{2k^{2}\lambda l^{2}qe}{\left( qe r_{h}+kl^{2}\right)^{3}  }+\frac{k^{2}\lambda l^{4}}{3r_{h}^{2}(kl^{2}+qer_{h})^{2}}-\frac{k^{2}\lambda qer_{h}^3}{2(kl^2+qer_{h})^{3}(kl^{2}+kr_{h}^{2}+qer_{h})}\right.  \right.
\\
\left. - \frac{\arctan\left(\frac{kr_{h}}{\sqrt{(kl^2+qer_{h})k}}\right)k^{2}\lambda qer_{h}(r_{h}-2kl^2)}{(qer_{h}^3+kl^2)^{3}\sqrt{(qer_{h}^3+kl^2)k}}
\pm\left(\frac{3}{4r_{h}^{4}}+\frac{2(qe-Mk)}{5kr_{h}^{5}}\right. \right.  \\
\left.  \left. \left. +\frac{k(M^2+q^2+4l^2)+2qe}{12r_{h}^{6}}-\frac{qe(M+3qek)}{7kr_{h}^{7}} -\frac{qe}{8r_{h}^{8}}\left(\frac{3l^2}{k}+\frac{M^2+q^2}{2k}\right)\right) \right] \right). \label{isq42}
\end{multline}
The influence of the NUT parameter $l$ on the GFs is more prominent, comparing with the bosons, for the fermions as can be deduced from Fig. \eqref{Figure5}. Increasing NUT parameter $l$ also increases the GFs, thus allow more fermionic thermal radiation to reach the observer at spatial infinity.

\section{QNM\lowercase{s} of CTNBH} \label{sec4}
QNM can be used to describe how BHs react to external perturbation. Today, there are various approaches to compute the QMMs (the readers can refer to \cite{SK1,SK2,SK3,SK4,SK5} and references therein), however, in this section, we shall consider the sixth-order WKB method \cite{SK6,SK7} to determine the QNMs of CTNBH for both fermionic and bosonic particles. The main equation of frequencies is given by \cite{SKL7,SKL8}, 
\begin{equation}
\omega^{2}=\left[  V_{0}+\sqrt{-2V_{0}^{\prime\prime}}\Lambda\left(  n\right)  -i\alpha \sqrt{-2V_{0}^{\prime\prime }}\left(  1+\Omega\left(  n\right)  \right)  \right]  ,\label{isq43}%
\end{equation}
where%
\begin{equation}
\Lambda\left(  n\right)  =\frac{1}{\sqrt{-2V_{0}^{\prime\prime
}}}\left[  \frac{1}{8}\left(  \frac{V_{0}^{\left(  4\right)  }}{V_{0}^{\prime\prime}}\right)  \left(  \frac{1}{4}+\alpha^{2}\right)  -\frac{1}{288}\left(
\frac{V_{0}^{\prime\prime\prime}}{V_{0}^{\prime\prime}}\right)  ^{2}\left(  7+60\alpha^{2}\right)  \right]  ,\label{isq44}%
\end{equation}
and%

\begin{multline}
\Omega\left(  n\right)  =\frac{1}{-2V_{0}^{\prime\prime}}\left[  \frac{5}{6912}\left(  \frac{V_{0}^{\prime\prime\prime}}{V_{0}^{\prime\prime
}}\right)  ^{4}\left(  77+188\alpha^{2}\right)  -\frac{1}{384}\left(
\frac{V_{0}^{\prime\prime\prime 2}V_{0}^{\left(  4\right)  }}{V_{0}^{\prime\prime3}}\right)  \left(  51+100\alpha^{2}\right)  +\right.  \\
\left.  \frac{1}{2304}\left(  \frac{V_{0}^{\left(  4\right)  }}{V_{0}^{\prime\prime}}\right)  ^{2}\left(  67+68\alpha^{2}\right)  +\frac{1}{288}\left(
\frac{V_{0}^{\prime\prime\prime}V_{0}^{\left(  5\right)  }}{V_{0}^{\prime\prime 2}}\right)  \left(  19+28\alpha^{2}\right)  -\frac{1}{288}\left(  \frac
{V_{0}^{\left(  6\right)  }}{V_{0}^{\prime\prime
}}\right)  \left(  5+4\alpha^{2}\right)  \right], \label{isq45}
\end{multline}
in which  
\begin{equation}
\alpha=n+\frac{1}{2},    \qquad V_{0}^{(n)}=\frac{d^{n}V(r^0_{\ast})}{dr_{\ast}^{n}}. \label{isq46}
\end{equation}

 The value of $r^0_{\ast}$ seen in Eq. \eqref{isq46} is the value where the effective potential reaches at its maximum.
 
 \subsection{Scalar QNM\lowercase{s}}
In this part, to explore the scalar QNMs of the CTNBH, we first consider the bosonic effective potential \eqref{isq36} within the QNM expression \eqref{isq43}. The effect of the NUT parameter $l$ on the QNMs is investigated. To this end, two sets of charge values are used: $q=\tilde{q}=0.1$ and $q=\tilde{q}=0.2$ for $n=0, n=1$, and $l=1$ states. The results are tabulated in Table \ref{tab1}, which shows that increase in the NUT parameter $l$ value, in general, decreases both the oscillation (real frequencies) and damping (imaginary frequencies). Moreover, for $n=1$ state although the real part remains intact, but the damping modes raises until the NUT parameter $l=3.3$ then they start to decrease for the both sets of the charges, $q=\tilde{q}=0.1$ and $q=\tilde{q}=0.2$. \\

%\begin{minipage}[c]{.40\textwidth}
\begin{table}[H]
   \centering
   \begin{tabular}{ |c|c|c|c|c|c|c|c|}
\hline
$\tilde{l}$ & $n$ & $q=\tilde{q}$ & $l$ & $\omega_{Bosons}$ & $q=\tilde{q}$& $l$ & $\omega_{Bosons}$\\
\hline\hline
1 & 0 & 0.2 & 3 & 0.2725502408-0.6131029148i & 0.1 & 3 & 0.2756084514-0.6136558658i\\
  &   &  & 3.1 & 0.2560591000-0.6082054931i& & 3.1 & 0.2587389727-0.6087803472i \\
  &   &  & 3.2 & 0.2409607665-0.6025973599i & & 3.2 & 0.2433163930-0.6031804616i\\
  &  & & 3.3 & 0.2271104777-0.5964385698i & & 3.3 & 0.2291872658-0.5970200571i \\
  &   & & 3.4 & 0.2143806411-0.5898593359i & & 3.4 & 0.2162166481-0.5904320921i \\
  &   & & 3.5 & 0.2026583779-0.5829655802i& & 3.5 & 0.2042858309-0.5835246352i \\
\hline
 & 1 & 0.2 & 3 & 2.248092173-1.900912954i& 0.1& 3 & 2.268320069-1.905126791i \\
  &   &  & 3.1 & 2.125469414-1.913465607i& & 3.1 & 2.143321140-1.917860042i\\
  &   & & 3.2 & 2.012248433-1.920073137i& & 3.2 & 2.028053175-1.924547860i \\
  &  & & 3.3 & 1.907546817-1.921767172i& & 3.3 & 1.921581800-1.926248930i\\
  &   & & 3.4 & 1.810574732-1.919400788i& & 3.4 & 1.823073953-1.923836247i\\
  &   & & 3.5 & 1.720624552-1.913681044i& & 3.5 & 1.731786697-1.918032284i \\
\hline
\end{tabular}
 \captionof{table}{Bosonic QNMs of the CTNBH for various combinations of charge and state values.} \label{tab1}
 \end{table}
%\end{minipage}

\begin{figure}[H]
\centering
\includegraphics[scale=.5]{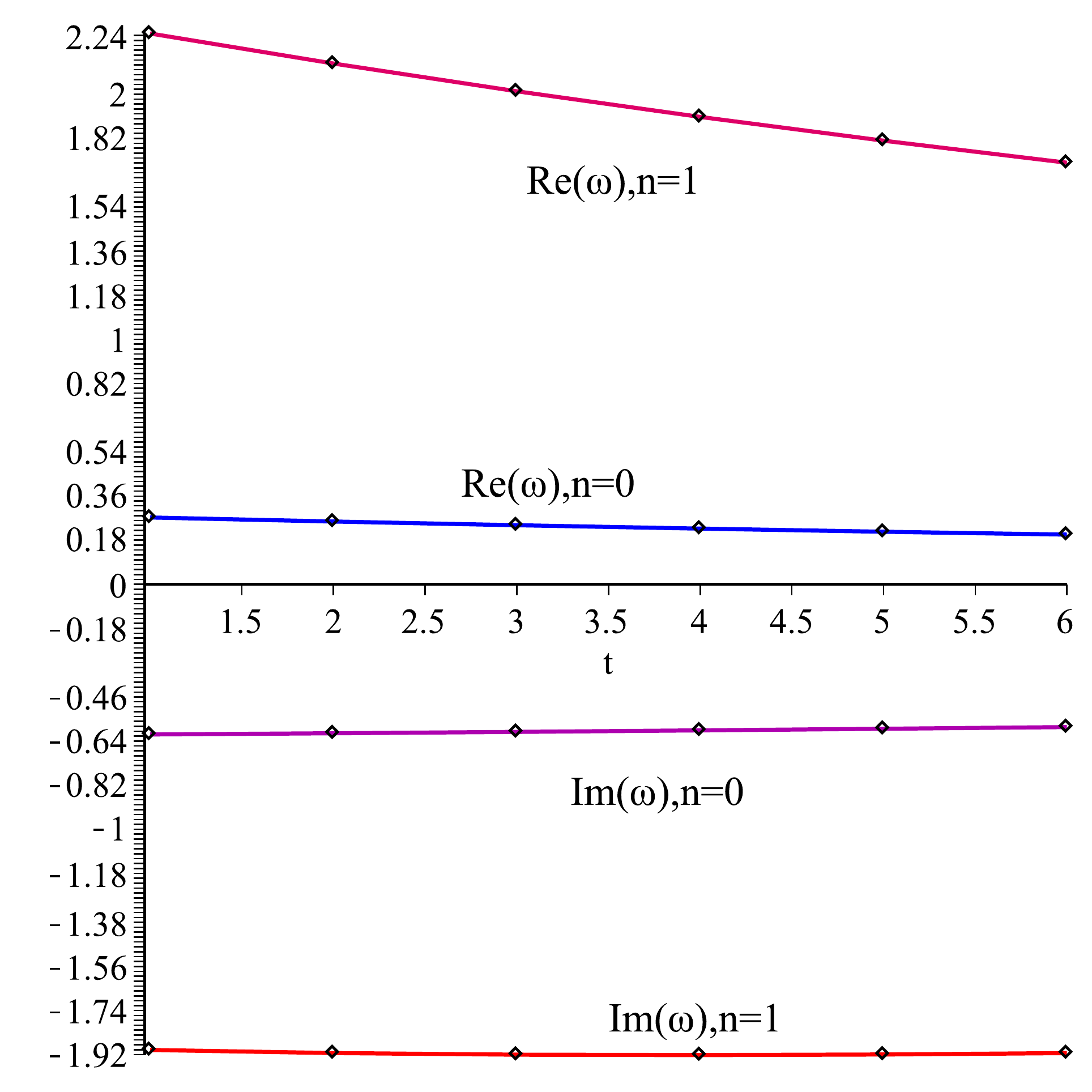} \caption{Graph of scalar QNMs for various values of the NUT parameter $l$ based on Table \ref{tab1}. Here, the considered charge values are $q=\tilde{q}=0.2$
.}
\label{Figure6}%
\end{figure}

To make the behaviors of bosonic QNMs more understandable, we have plotted Fig. \ref{Figure6} for varying NUT parameter $l$ and for different values of charges $q=\tilde{q}=0.2$. It is also worth noting for other low charge values like $q=\tilde{q}=0.1$, the figures plotted are almost identical to Fig. \ref{Figure6} are obtained \big(that is why we have only depicted Fig. \ref{Figure6}\big).

\subsection{Dirac QNM\lowercase{s}}
For computing the Dirac QNMs in the CTNBH spacetime, we simply insert the Dirac effective potential \eqref{isq28} into the GF expression \eqref{isq43}. After computing the frequencies,  we have tabulated them in Table \ref{tab2}, which presents the Dirac QNMs.  Table \ref{tab2} results are obtained from  $k=0.1$ and $M=1$ (for the other pairwise values of $k$ and $M$, one gets similar behaviors). The results indicate that while the fermionic oscillation frequencies increase smoothly with the increasing NUT parameters $l$ for both $n=0$ and $n=1$ values, but this impression becomes inverse for the damping modes as happened in the bosonic case. On the other hand, it is worthy to mention that the evolution (in the context of increase and decrease) of the QNM frequencies of the Dirac particles with the NUT parameter is in harmony with their GFs. 

%\begin{minipage}[c]{.40\textwidth}
\begin{table}[H]
   \centering
   \begin{tabular}{ |c|c|c|c|c|c|c|c|}
\hline
$\tilde{l}$ & $n$ & $q=e$ & $l$ & $\omega_{Fermions}$ & $q=e$& $l$ & $\omega_{Fermions}$\\
\hline\hline
1 & 0 & 0.9 & 1 & 0.2246744266-0.1507449260i & 0.8 & 1 & 0.2724250498-0.1725874216i\\
  &   &  & 1.1 & 0.2251419577-0.1502503846i& & 1.1 & 0.2731628029-0.1716878385i \\
  &   &  & 1.2 & 0.225654448-0.1497044313i & & 1.2 & 0.2739749307-0.1706919821i\\
  &  & & 1.3 & 0.2262187771-0.1491057793i & & 1.3 & 0.2748624250-0.1695967016i \\
  &   & & 1.4 & 0.2268297702-0.1484531124i & & 1.4 & 0.2758266950-0.1683984787i \\
  &   & & 1.5 & 0.2274901744-0.1477447054i& & 1.5 & 0.2768689571-0.1670932364i \\
\hline
 & 1 & 0.9 & 1 & 1.456007708-0.7655868096i& 0.8 & 1 & 1.763833077-0.8187154684i \\
  &   &  & 1.1 & 1.459193911-0.7592922722i& & 1.1 & 1.768465908-0.8073416560i\\
  &   & & 1.2 & 1.462685227-0.7523492681i& & 1.2 & 1.773520512-0.7947675850i \\
  &  & & 1.3 & 1.466482391-0.7447418782i& & 1.3 & 1.778990042-0.7809589163i\\
  &   & & 1.4 & 1.470585762-0.7364546462i& & 1.4 & 1.784867095-0.7658771666i\\
  &   & & 1.5 & 1.474995402-0.7274692890i& & 1.5 & 1.791142115-0.7494791876i \\
\hline
\end{tabular}
   \captionof{table}{Fermionic QNMs of the CTNBH for various combinations of charge and state values.} \label{tab2}
\end{table}
%\end{minipage}

\begin{figure}[h]
\centering
\includegraphics[scale=.5]{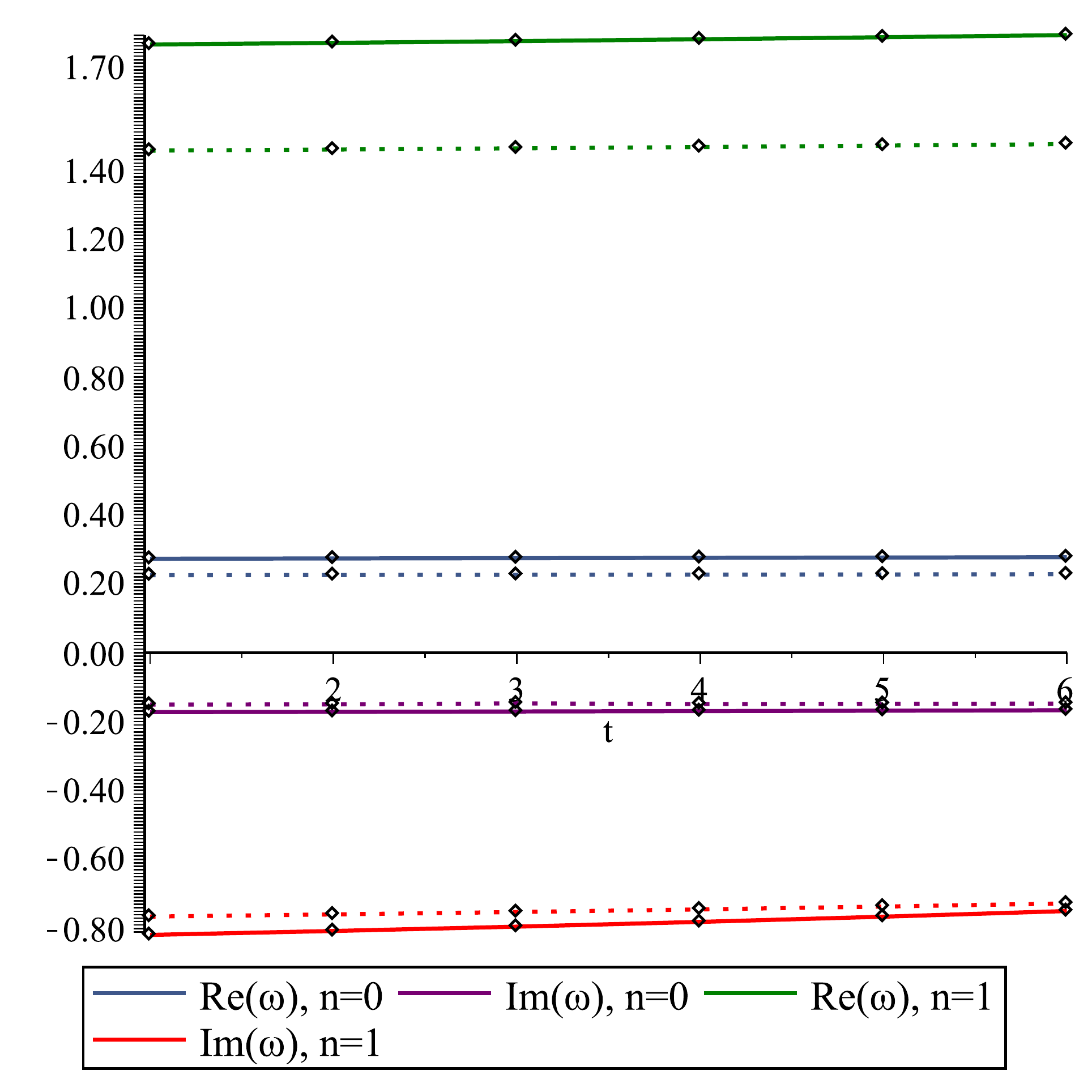} \caption{Graph of Dirac QNMs 
for various values of the NUT parameter $l$ based on Table \ref{tab2}. As the solid lines are for $q=e=0.8$, the dotted lines stand for $q=e=0.9$.}
\label{Figure8}%
\end{figure}

To understand the behaviors of imaginary and real frequencies,  we have illustrated the Dirac QNMs in Fig. \ref{Figure8}. The both behaviors are in accordance not only with each other, but with the bosonic QNMs as well.

\section{Conclusion} \label{sec5}
In this study, we have thoroughly investigated the spinorial wave equations in the background of the CTNBH. Since the metric of CTNBH describes the vacuum spacetime around a source having three physical parameters -- mass, charge, and NUT parameter -- we, in particular, have aimed to reveal the effect of the NUT parameter on the CTNBH's thermal radiation. To this end, we have considered the charged fermion and scalar perturbations of the CTNBH. The Dirac equation has been separated into radial and angular parts by means of a null tetrad of the NP formalism. The effective potentials of the fermions and bosons for the CTNBH geometry have been derived. We have also shown (see Figs. \ref{fig1} and \ref{Figure3}) that the shapes of the effective potentials strictly depend on the values of the NUT parameter $l$. The plots indicate that potential barriers' peaks are higher for the small values of the NUT parameter which means that it is more difficult for the thermal radiation to reach the distant observer. The latter remark naturally show itself in the GFs. We have shown with Figs. \ref{Figure4} and \ref{Figure5} that GFs decrease at low NUT values and vice versa.

As for the QNMs of the CTNBH, we have obtained the following important results for the bosonic and fermionic QNMs: (i) increasing NUT parameter decreases both real and imaginary values of the bosonic QNM frequencies (ii) while the imaginary part of the fermionic QNM frequencies decreases with the increasing NUT parameter, however the real part increases, which means more frequent oscillations, unlike the general convention. In summary, with increasing NUT parameter, bosonic and fermionic QNMs will be damped earlier. The results obtained are tabulated in Tables \ref{tab1} and \ref{tab2} and graphically visualized with Figs. \ref{Figure6} and \ref{Figure8}. 

In the near future, we would like to extend this work to a stationary CTNBH. By doing this, we want to consider not only the rotation, but also the modified gravitational theories that try to better model our Universe. In this regard, we plan to work on charged and rotating NUT BHs in Rastall gravity \cite{Prihadi:2019isb}, which might give us the effect of dark matter and dark energy on BH's thermal radiation, in addition to the effects of rotation and NUT parameter. We believe that the results to be obtained will be interesting.

\bigskip
\bigskip
\begin{quote}
{\LARGE Appendix A}\label{apA}
\end{quote}

 In the NP formalism \cite{is17}, one can introduce the covariant 1-forms of the CTNBH metric (\ref{isq1}) as follows:

\[
l_{\mu }=\bigg(\frac{1}{f\left( r\right) },1,0,-2l\cos \theta \bigg), 
\]%
\[
n_{\mu }=\bigg(\frac{f\left( r\right) }{2},\frac{1}{2},0,-2lf(r)\cos \theta \bigg), 
\]%
\ 
\[
m_{\mu }=\sqrt{\frac{r^{2}+l^{2}}{2}}\bigg(0,0,-1,-i\sin \theta \bigg), 
\]%
\begin{equation}
\overline{m}_{\mu }=\sqrt{\frac{r^{2}+l^{2}}{2}}\bigg(0,0,-1,i\sin \theta\bigg ),\tag{A1}  \label{A1N}
\end{equation}
where a bar over a quantity denotes complex conjugation. Whence, the dual co-tetrad become
\[
l^{\mu }=\bigg(\frac{1}{f},1,0,0\bigg), 
\]
\[
n^{\mu }=\bigg(\frac{1}{2},-\frac{f}{2},0,0\bigg), 
\]
\[
m^{\mu }=\frac{1}{\sqrt{2\left( r^{2}+l^{2}\right) }}\bigg(2il\cot \theta ,0,1,%
\frac{i}{\sin \theta }\bigg), 
\]%
\begin{equation}
\overline{m}^{\mu }=\frac{1}{\sqrt{2\left( r^{2}+l^{2}\right) }}\bigg(-2il\cot
\theta ,0,1,\frac{-i}{\sin \theta }\bigg).\tag{A2}  \label{A2N}
\end{equation}
Using the above null tetrad \eqref{A1N}, one can compute the non-zero of the spin coefficients \cite{is17} of the CTNBH as follows
\[
\rho =\frac{-r+il}{r^{2}+l^{2}};\qquad \mu =\frac{-r+il}{r^{2}+l^{2}}\left( 
\frac{f}{2}\right) , 
\]
\[
\epsilon =\frac{il}{2\left( r^{2}+l^{2}\right) };\qquad \gamma =\frac{1}{4}%
\left( f^{\prime }(r)+\frac{ilf}{r^{2}+l^{2}}\right) , 
\]
\begin{equation}
\alpha =\frac{-\cot \theta }{2\sqrt{2\left( r^{2}+l^{2}\right) }};\qquad
\beta =\frac{\cot \theta }{2\sqrt{2\left( r^{2}+l^{2}\right) }}.\tag{A3}  \label{A3N}
\end{equation}%
\bigskip Finally, the corresponding directional derivatives become
\[
D=l^{\mu }\partial _{\mu }=\frac{1}{f}\partial t+\frac{\partial }{%
\partial r},\qquad 
\]%
\[
\Delta =n^{\mu }\partial _{\mu }=\frac{1}{2}\frac{\partial }{\partial t}-%
\frac{f}{2}\frac{\partial }{\partial r},
\]%
\[
\delta =m^{\mu }\partial _{\mu }=\frac{1}{\sqrt{2\left( r^{2}+l^{2}\right) }}%
(2il\cot \theta \frac{\partial }{\partial t}+\frac{\partial }{\partial
\theta }+\frac{i}{\sin \theta }\frac{\partial }{\partial \phi }),
\]%
\begin{equation}
\bar{\delta}=\overline{m}^{\mu }\partial _{\mu }=\frac{1}{\sqrt{2\left(
r^{2}+l^{2}\right) }}(-2il\cot \theta \frac{\partial }{\partial t}+\frac{%
\partial }{\partial \theta }-\frac{i}{\sin \theta }\frac{\partial }{\partial
\phi }).\tag{A4}  \label{A4N}
\end{equation}

\end{document}